\documentclass{article}
\usepackage{latticelab,times}
\nonanonymous

\usepackage[utf8]{inputenc} 
\usepackage[T1]{fontenc}    
\usepackage{hyperref}       
\usepackage{url}            
\usepackage{booktabs}       
\usepackage{amsfonts}       
\usepackage{nicefrac}       
\usepackage{microtype}      
\usepackage{xcolor}         

\usepackage{amssymb}
\usepackage{amsmath}
\usepackage{amsthm}
\usepackage{tcolorbox}
\usepackage{siunitx}
\theoremstyle{plain}

\theoremstyle{definition}

\usepackage[acronym]{glossaries} 
\usepackage{pifont}              
\usepackage{paralist}            
\usepackage{comment}             
\usepackage{mathtools}           
\usepackage{multirow}            
\usepackage{caption}             
\usepackage{minitoc}             
\usepackage{wrapfig}             
\usepackage{subcaption}
\usepackage[percent]{overpic}
\usepackage{booktabs}
\usepackage{tabularx}
\usepackage{array}
\usepackage{makecell}
\usepackage{xcolor} 
\definecolor{LightGray}{gray}{0.9}
\newcommand{\bse}{\begin{subequations}}
\newcommand{\ese}{\end{subequations}}

\usepackage{todonotes}

\newif\ifrevised
\revisedtrue    

\ifrevised
  
\else
  
\fi

\title{LLM-Guided Test-Time Discovery of Quantum-Chemical Approximation Algorithms}

\paperlogo{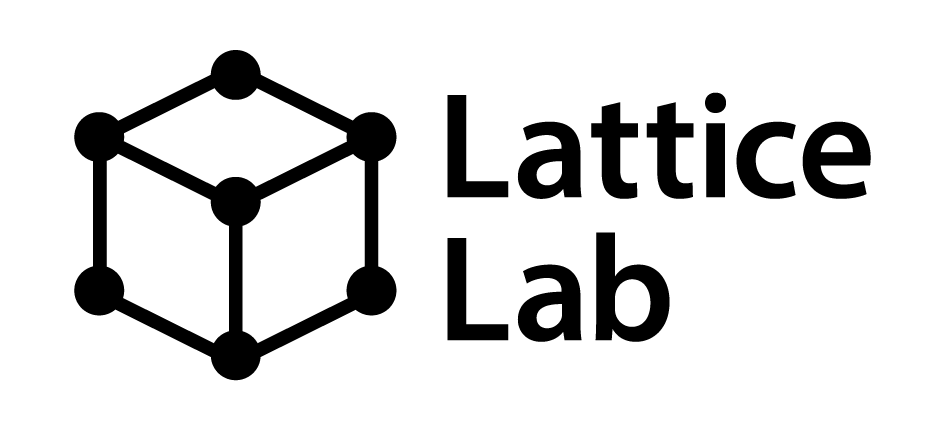}

\author{
    Masaya Hagai$^{1,2}$, 
    Yuta Suzuki$^{1}$,
    Tomoya Murata$^{1}$,
    Shuhei Kurita$^{1,3}$,
    Masaki Adachi$^{1}$,\\
    \small{$^1$ Lattice Lab, Toyota Motor Corporation}\\
    \small{$^2$ Department of Chemistry, Graduate School of Science, Nagoya University}\\
    \small{$^3$ National Institute of Informatics}\\
    \small{\textit{M.H. performed this work during an internship at Lattice Lab, Toyota Motor Corporation.}}\\
    \small{ \texttt{\url{yuta_suzuki_ah@mail.toyota.co.jp}}
    }\\
}

\begin{document}

\maketitle

\begin{abstract}
Quantum chemistry simulations underpin modern materials discovery, yet their impact is limited by steep computational cost and dependence on fixed approximation schemes. Foundation models, such as machine-learned interatomic potentials, have accelerated parts of this workflow, but their reliance on large-scale pretraining restricts adaptability at the frontier of chemical space, where methodological innovation and sparse data are the norm. Agentic AI systems can automate existing simulation pipelines, yet they remain constrained by the predefined tools and algorithms they orchestrate.
In response, we introduce LADeQ, an LLM-guided workflow that discovers, implements, and benchmarks candidate approximation algorithms at test-time within existing quantum chemistry codes. Rather than selecting from a predefined repertoire, LADeQ constructs candidate approximation schemes on demand, drawing on techniques from disciplines such as spatial statistics, circuit simulation, and kernel methods that have had little prior presence in electronic-structure theory. Because it builds on an out-of-the-box language model, LADeQ requires no task-specific pretraining or curated data, and the resulting implementations are transparent and inspectable, with explicitly traceable approximation errors that enable principled control of accuracy--efficiency trade-offs.
We show that LADeQ accelerates coupled cluster singles and doubles (CCSD) and configuration interaction singles and doubles (CISD) calculations while keeping correlation-energy errors within user-specified tolerances, demonstrating autonomous, objective-driven discovery of approximation algorithms inside existing electronic-structure solvers.
\end{abstract}

\section{Introduction}\label{sec:intro}
Recent advances in generative models for molecules, drugs, and crystalline materials have enabled the in silico generation of vast numbers of candidate structures for materials discovery \cite{Gangwal2024-uj,Zeni2025-pq,Lin2025-gt}. Combined with large language models (LLMs) \cite{Ramos2025-hf,Gruver2024-rg} and agentic AI systems \cite{Ramos2025-hf,Zou2025-en}, these approaches increasingly automate the end-to-end materials design workflow--from idea generation to database querying and experiment planning--lowering the barrier to entry for non-experts and democratizing discovery processes that previously required years of domain-specific training.

However, the generation of candidate materials and automation of workflows alone does not guarantee the discovery of useful materials. Utility is ultimately determined by target properties, whose reliable evaluation typically relies on quantum chemical simulations. These simulations remain a dominant computational bottleneck in materials discovery pipelines due to their high cost, limiting the scale and speed of exploration. Consequently, practical discovery hinges on approximation algorithms that balance accuracy and efficiency. The choice of approximation method has a decisive impact on the plausibility and reliability of predicted materials.

To alleviate this bottleneck, growing attention has turned to simulation foundation models, particularly machine-learning interatomic potentials (MLIPs)\cite{Behler2007-pf,Kocer2022-bf}, which learn surrogate representations of quantum chemical simulations from large-scale datasets. MLIPs enable rapid inference and have proven effective for large systems, long time scales, and many-body simulations \cite{Unke2024-fb,Zhang2024-ta,Zhang2025-nf}, substantially accelerating materials screening workflows. 
Despite these successes, MLIPs face two fundamental limitations: reliability outside their validated domain and flexibility across levels of theory.
Although uncertainty estimation and active-learning strategies have improved reliability, certifying predictions for chemically or structurally novel candidates remains challenging, and high-level quantum chemistry simulations are often still required for final validation \cite{Gangwal2024-uj}.
Moreover, because MLIP datasets are usually generated at a single, fixed level of theory to ensure consistency \cite{Unke2021-zh}, adapting them to alternative electronic-structure methods generally requires new reference data, retraining, and validation.

Density functional theory (DFT) \cite{Kohn1965-mi} is widely used for dataset construction due to its favorable cost–accuracy trade-off and remains indispensable for large-scale computational screening. At the same time, high-level wavefunction-based methods play an important role as reference calculations for benchmarking, validation, and methodological development. Their computational cost grows steeply with system size, motivating algorithmic and numerical approximations that reduce runtime while retaining controlled accuracy. Incorporating multiple quantum chemistry methods into training datasets introduces heterogeneity that degrades predictive performance \citep{muandet2022impossibility}, leading to fragmented datasets and models. In such settings, more adaptive and flexible computational strategies are needed.

In parallel, agentic AI systems have emerged to automate scientific workflows. Systems such as ChemCrow \cite{M-Bran2024-cz}, Coscientist \cite{Boiko2023-xs}, and LLMatDesign \cite{Jia2024-eq} assist in synthesis planning, candidate generation, and database querying, while others automate simulation setup, execution, and analysis (e.g., MDAgent\cite{Breazeal2024-yv}, DynaMate\cite{Mendible-Barreto2025-co}, MDCrow\cite{Campbell2025-ds}, El Agente \cite{Zou2025-en}). Although these approaches greatly improve workflow efficiency, they treat simulators as black-box tools and do not modify the underlying computational algorithms. As a result, the core bottleneck--the simulator itself--remains unchanged, limiting overall acceleration.

We therefore ask whether LLM agents can move beyond workflow automation and contribute directly to approximation design inside electronic-structure solvers. We introduce \emph{LADeQ} (LLM-guided Automatic approximation Design for Quantum chemistry), an LLM-guided workflow that discovers, implements, and benchmarks candidate approximation algorithms at test-time within existing quantum chemistry codes. Rather than selecting existing simulators as black boxes, LADeQ dynamically constructs new approximation algorithms tailored to the computational objective, directly targeting the simulation bottleneck. The agent requires no pretraining, dataset construction, or model maintenance; instead, it leverages general reasoning capabilities to synthesize algorithmic ideas at runtime.

\begin{figure}[htbp]
  \centering
  \includegraphics[width=1.0\linewidth]{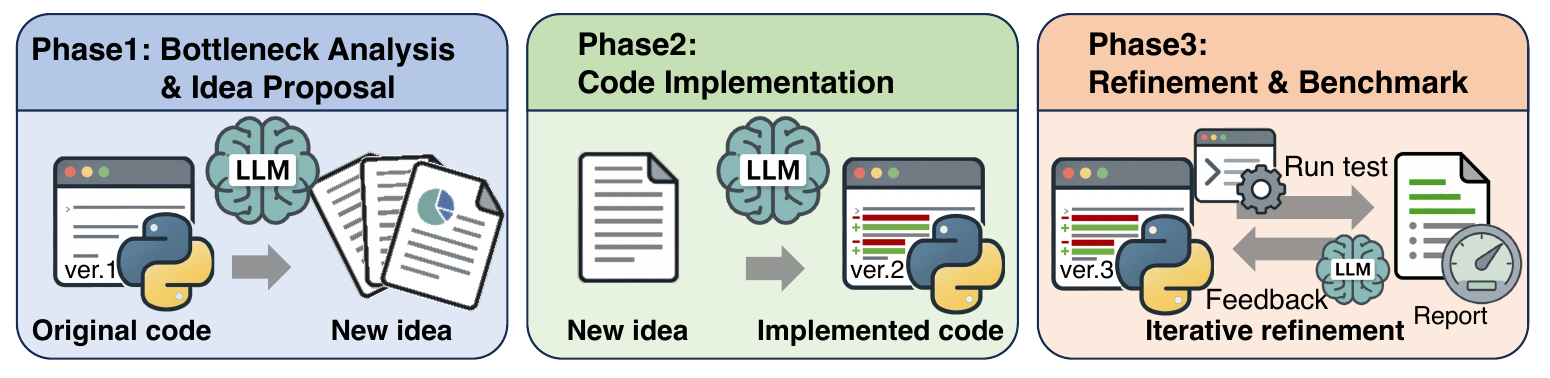}
  \caption{Overview of the LADeQ workflow. LADeQ operates in three phases.
In Phase 1, an LLM analyzes an existing electronic-structure code, identifies major computational bottlenecks, and proposes candidate approximation ideas. 
In Phase 2, each idea is translated into multiple concrete implementations for PySCF.
In Phase 3, the implementations are benchmarked using correlation-energy errors, wall-clock times, and runtime behavior traces; this feedback is used for iterative refinement and selection of effective ideas under prescribed accuracy constraints.}
  \label{fig:overall_flowchart}
\end{figure}

Notably, LADeQ’s sources of inspiration extend beyond quantum chemistry. 
The agent draws acceleration strategies from diverse scientific domains--including spatial statistics, image compression, and fluid dynamics simulation--that have been rarely explored in electronic-structure theory. 
Figure~\ref{fig:overall_flowchart} illustrates the LADeQ workflow: the agent analyzes existing electronic-structure code, identifies dominant computational bottlenecks, proposes candidate algorithmic modifications, and iteratively refines them until a performant solution is obtained. 
Crucially, all discovered algorithms are fully transparent, with explicitly traceable approximation errors, in contrast to black-box MLIP approaches. This transparency supports informed assessment of the accuracy--efficiency trade-off in the resulting calculations.

We view LADeQ as an initial step toward agent-driven algorithm discovery in quantum chemistry--one that not only accelerates simulations by adapting state-of-the-art methods on demand, but also serves as a source of inspiration for human researchers seeking new algorithmic pathways to push the limits of quantum chemical computation.

\section{Results}

We tested the central hypothesis that LADeQ can generate implementable approximation strategies for correlated wavefunction methods, reducing computational cost while controlling the error introduced into the correlation energy. We further examined whether the generated strategies depended on the numerical structure of the target method, rather than reflecting a single method-independent heuristic.

To this end, we selected two widely used but computationally demanding wavefunction methods: coupled-cluster singles and doubles (CCSD) and configuration interaction singles and doubles (CISD), using their PySCF\cite{PySCF} implementations as reference baselines. These methods represent distinct numerical regimes: CCSD involves nonlinear iterative amplitude equations, whereas CISD is formulated as a large linear eigenvalue problem.

All approximation-development experiments were conducted on the linear hydrocarbon C$_{15}$H$_{32}$ with the STO-3G basis set. For each target method, LADeQ first generated 20 approximation ideas. For each idea, we then ran 10 independent implementation trials, each starting from the same baseline code and the same idea description but following an independent LLM generation and refinement trajectory. Within each trial, the implementation was iteratively refined using execution logs, timing measurements, and correlation-energy errors, up to a maximum of 10 refinement iterations. Thus, an implementation trial denotes one independent attempt to realize a given idea in code, whereas a refinement iteration denotes one code-update step within that trial. This protocol yielded 200 implementation trials per target method.

To examine whether the selected approximations generalize beyond the development molecule, we benchmarked the best-performing candidates on additional molecules spanning molecular-size and electronic-structure variations. C$_{20}$H$_{42}$ was chosen as a controlled size-transfer test within the same saturated hydrocarbon family as C$_{15}$H$_{32}$, whereas coronene C$_{24}$H$_{12}$ and porphyrin C$_{20}$H$_{14}$N$_4$ were chosen as electronically distinct $\pi$-conjugated systems. This benchmark design allows us to distinguish approximations that transfer within a chemically similar family from those that remain effective across different electronic delocalization patterns.

Each implementation was benchmarked against the baseline in terms of wall-clock time and correlation-energy relative error. In addition, we compared the discovered approximations with the domain-based local pair natural orbital (DLPNO) method--an established, expert-designed acceleration framework based on a fixed locality principle--to assess whether LADeQ can construct distinct, tailored strategies that achieve competitive or improved runtime–accuracy trade-offs relative to a predefined, general-purpose approach.

\subsection{Quantitative analysis}
\begin{figure}[htbp]
  \centering
  \includegraphics[width=1.0\linewidth]{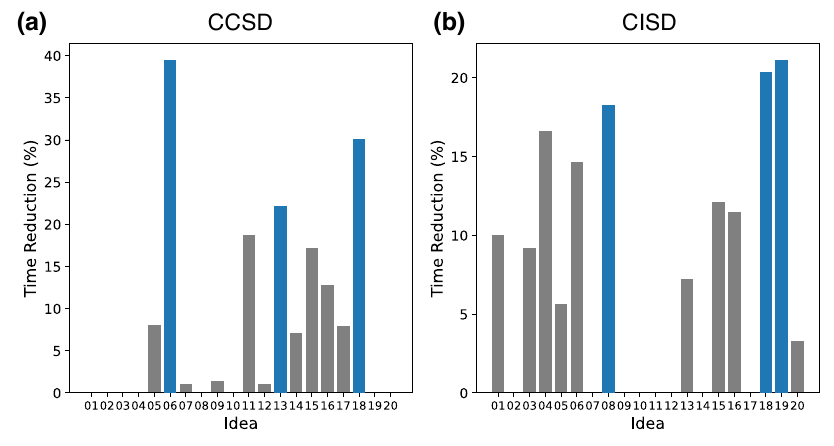}
  \caption{Runtime reduction achieved by the best implementation trial for each approximation idea: (a) CCSD and (b) CISD. 
For each idea, the plotted value corresponds to the largest runtime reduction among its 10 independent implementation trials that satisfied the error criterion. 
Blue bars indicate the top three ideas for each method. A value of 0 indicates that no implementation trial for that idea achieved both reduced runtime and acceptable correlation-energy error.}
  \label{fig:best_time_reduction}
\end{figure}

We first assessed LADeQ's ability to generate successful approximation implementations on the development molecule C$_{15}$H$_{32}$. 
To this end, we quantified (i) the success rate of each LLM-proposed idea, defined as the fraction of its implementations that achieved a runtime reduction of $\geq 2\%$ relative to the baseline while maintaining a correlation-energy error $\le 10\%$, and (ii) the magnitude of runtime reduction achieved by the successful implementations. 
At the idea level, 12 of the 20 proposed ideas for each method yielded at least one implementation trial satisfying the predefined runtime--accuracy thresholds, corresponding to an idea-level success rate of 60\% for both CCSD and CISD.
This indicates that a majority of independently generated approximation concepts can be translated into effective, accuracy-preserving accelerations.
Figure~\ref{fig:best_time_reduction} summarizes the maximum runtime reduction achieved by each idea under the error constraint. For each idea, the bar height represents the best-performing implementation among its 10 independent realizations. A zero-height bar indicates that no implementation simultaneously satisfied the accuracy requirement and reduced runtime. The top three ideas for each method are highlighted in blue.
For CCSD, the largest runtime reduction reached 39.4\% (idea\_06), while for CISD the best-performing idea (idea\_19) achieved a 22\% reduction. Importantly, multiple distinct ideas produced substantial speedups for both methods, demonstrating that LADeQ does not rely on a single dominant heuristic but instead identifies diverse, method-dependent acceleration strategies. Together, these results support the hypothesis that LADeQ can autonomously generate effective approximation algorithms that improve efficiency while preserving accuracy.

\begin{figure}[htbp]
  \centering
  \includegraphics[width=1.0\linewidth]{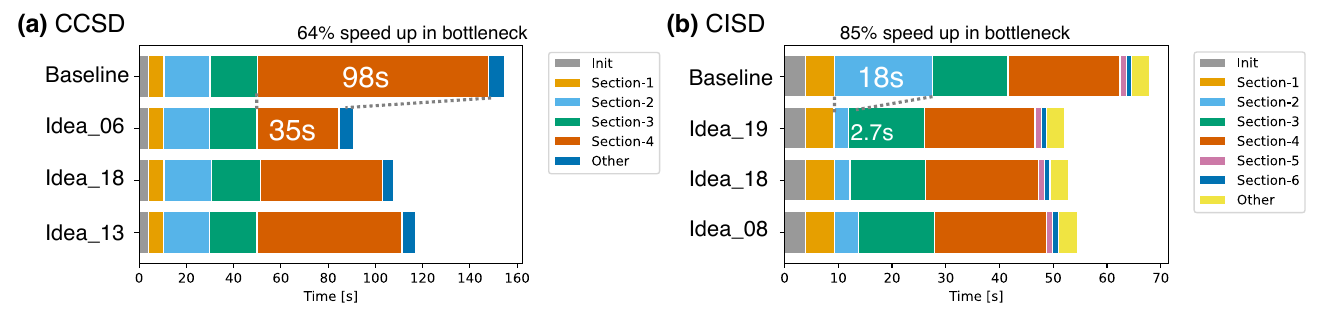}
  \caption{Section-wise runtime breakdown for the baseline and the top three implementations for (a) CCSD and (b) CISD, measured using PySCF’s built-in per-section timers. For readability, PySCF’s internal timer labels are relabeled as \texttt{Section-1} through \texttt{Section-N}. \texttt{Init} denotes initialization time and \texttt{Other} aggregates the remaining time. The white labels and dashed guide lines highlight the runtime reduction of the main affected section, from 98~s to 35~s for CCSD and from 18~s to 2.7~s for CISD.}
  \label{fig:bench_section}
\end{figure}

Next, we investigated whether the observed accelerations arose from targeted reductions in major runtime components rather than from diffuse or incidental changes across the codebase. 
To this end, we decomposed the total runtime into individual computational sections and examined which sections were shortened by the LLM-generated approximations. This analysis allows us to determine whether each approximation acts on a specific computational bottleneck or merely changes overall runtime in an uncontrolled manner.
Figure \ref{fig:bench_section} shows the runtime breakdown obtained from PySCF's built-in per-section timers for the baseline implementation and the top three implementations.
For clarity, PySCF's internal section names are relabeled as neutral identifiers, from \texttt{Section-1} to \texttt{Section-N}, while keeping \texttt{Init} and \texttt{Other} to denote initialization and remaining time.
For CCSD, all of the top three approximate implementations reduced the runtime of the largest baseline bottleneck, \texttt{Section-4}, while non-target sections remained nearly unchanged. In the fastest implementation, idea\_06, the time spent in \texttt{Section-4} decreased from 98~s in the baseline to 35~s, corresponding to a 64\% reduction.
For CISD, the largest runtime reduction occurred in \texttt{Section-2}, which was the second-largest baseline component but the main component affected by the LLM-generated modifications. In the fastest implementation, idea\_19, the runtime of \texttt{Section-2} was reduced by 85\%. Thus, the CISD speedup did not arise from uniformly shortening all parts of the calculation, but from selectively reducing a major reducible component of the solver.
Together, these results show that LADeQ-generated approximations act on identifiable runtime components in a method-dependent manner.

\begin{figure}[htbp]
  \centering
  \includegraphics[width=1.0\linewidth]{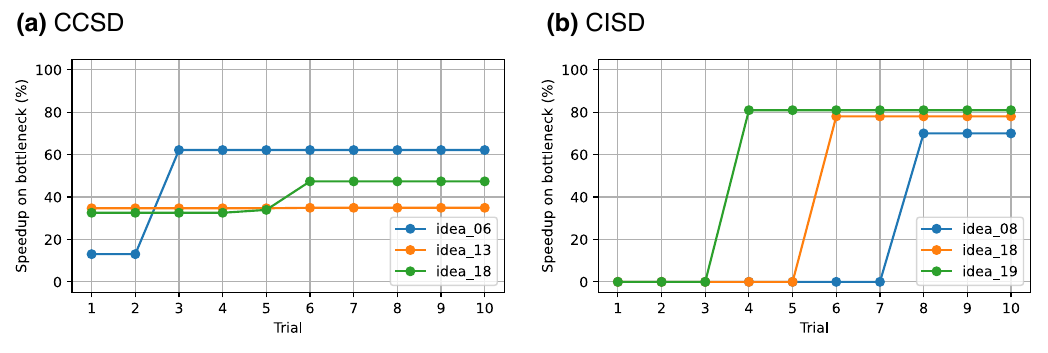}
  \caption{Best-so-far trajectories of runtime reduction over 10 independent implementation attempts for the top three approximation ideas for (a) CCSD and (b) CISD. 
  The metric is defined as the runtime reduction in the largest affected section: \texttt{Section-4} for CCSD and \texttt{Section-2} for CISD.}
  \label{fig:best_time_transition}
\end{figure}

Because independently generated implementations vary in both accuracy and speed, we next examined how rapidly effective implementations emerge as additional trials are sampled. 
Figure~\ref{fig:best_time_transition} shows the best-so-far runtime reduction over 10 independent implementation trials for the top three ideas in each method.
For CCSD, the reduction is measured in the primary bottleneck section, whereas for CISD it is measured in \texttt{Section-2}, the main section affected by the selected approximations.
For both methods, the best-so-far reduction largely saturates after approximately eight trials, suggesting that 10 independent trials per idea are sufficient for identifying strong runtime-reduction candidates under the present search protocol.
While the best-so-far runtime reduction saturates, additional attempts can still be beneficial for improving accuracy, especially when targeting stricter error thresholds.
Therefore, for the goal of maximizing runtime reduction under the present accuracy criterion, 10 implementation attempts per idea are sufficient and constitute a reasonable choice for the number of trials.
Since attempts are mutually independent and therefore fully parallelizable, the primary limiting factor is the API-cost budget. In our setting, executing one implementation attempt, consisting of code generation and iterative refinement, results in 7.4 LLM API calls on average and an average API cost of 1.1~USD per attempt.

\begin{figure}[htbp]
  \centering
  \includegraphics[width=1.0\linewidth]{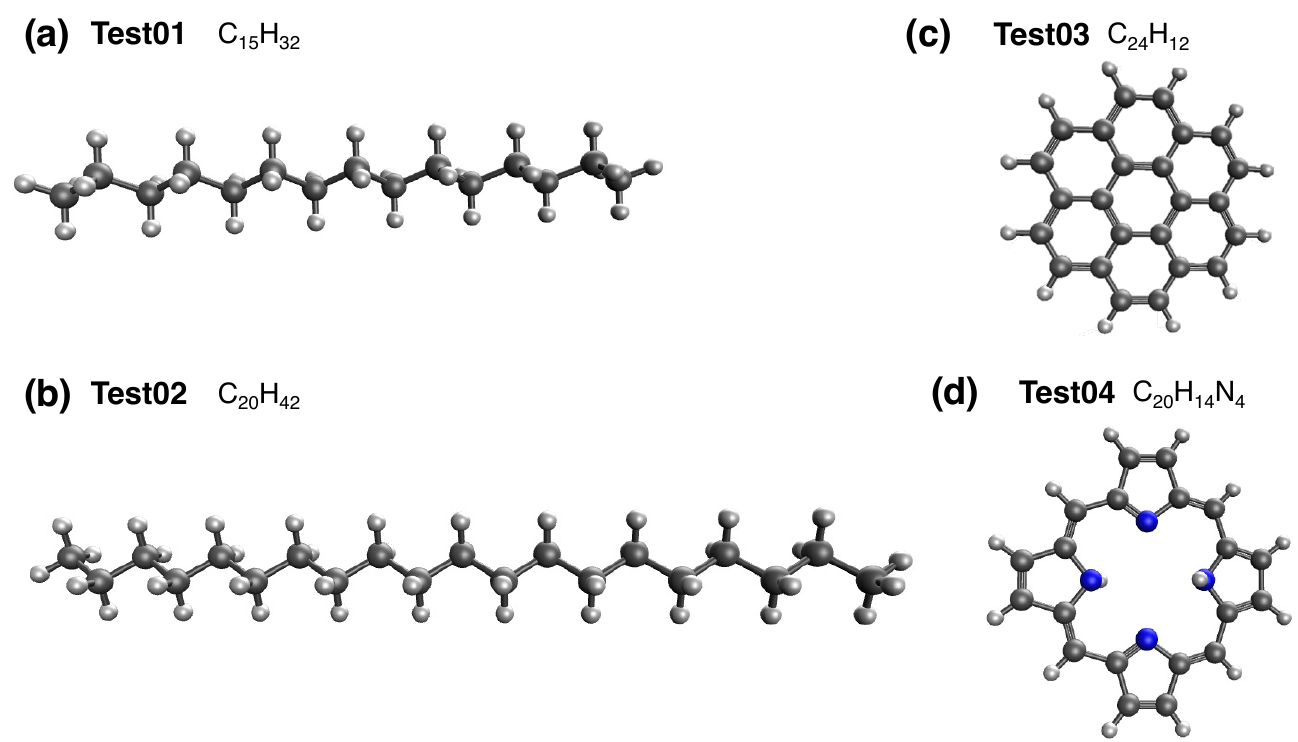}
  \caption{Molecules used in the benchmark: (a) linear hydrocarbon C$_{15}$H$_{32}$, (b) linear hydrocarbon C$_{20}$H$_{42}$, (c) coronene, and (d) porphyrin.}
  \label{fig:bench_mol}
\end{figure}

To evaluate the generality of the LLM-generated approximations across different molecules, we performed additional benchmark measurements using four molecules (Test01–04) shown in Figure \ref{fig:bench_mol}. 
Since the approximation design is guided by performance feedback on Test01, we need to assess whether the resulting approximation strategy generalizes beyond the development test case, rather than reflecting molecule-specific tuning.
To this end, we constructed a benchmark set that varies (i) molecular size within a fixed chemical class and (ii) electronic delocalization, which can change the cost profile and bottlenecks of correlated wavefunction calculations.
Test01 is the same linear hydrocarbon, C$_{15}$H$_{32}$, used as the benchmark molecule in the approximation-development process, and Test02 is the larger linear hydrocarbon C$_{20}$H$_{42}$.
This pair provides a controlled size-scaling test within a nearly identical chemical motif.
Test03 and Test04 are delocalized $\pi$-conjugated systems, represented by coronene and porphyrin, respectively. 

Compared with alkanes (Test01 and Test02), $\pi$-conjugated molecules differ markedly  in electronic delocalization and orbital structure, which may affect the performance of CISD/CCSD approximations that exploit locality and sparsity.
All benchmarks used the STO-3G basis set and were performed in a single-thread setting.

Based on the results described above, we selected three promising implementations each for CCSD and CISD and evaluated the runtime reduction and correlation-energy error for each test molecule (Table~\ref{tab:ccsd-cisd-speed-energy}).
For CCSD, the results reveal a clear contrast between approximations that appear selectively effective within the molecular family used during approximation development and those that remain effective across chemically distinct molecules in the benchmark set.
In particular, some implementations that were effective for the saturated alkane family represented by Test01–02 became less effective for the delocalized $\pi$-conjugated molecule (Test03).
For example, idea\_06 increased the runtime substantially for Test03 and Test04. 
In addition, idea\_13 failed to reach the CCSD convergence criterion within the default maximum of 50 iterations in the amplitude-update loop for Test03.
In contrast, idea\_18 achieved a comparable runtime reduction of approximately 30\% across all benchmark molecules, including the delocalized $\pi$ system Test03, while keeping the correlation energy error below 0.01\%.
This indicates that, although some LLM-generated ideas can be specialized, our approach can also yield approximation strategies that transfer robustly across distinct molecular classes and delocalization regimes.
For CISD, all selected implementations yielded runtime reductions for all test molecules. 
Among them, idea\_19 achieved an average runtime reduction of 23\% while maintaining the correlation-energy error within 0.01\%, representing the best overall trade-off within our tested LLM-generated CISD approximations.

\begin{table}[htbp]
  \centering
  \caption{Runtime reductions and relative correlation-energy errors for CCSD and CISD using different approximations across the four benchmark molecules. Negative runtime reduction indicates that the approximate calculation is slower than the calculation without the approximation. N/A indicate that the calculation did not converge within the default maximum of 50 iterations in the amplitude-update loop.}
  \label{tab:ccsd-cisd-speed-energy}
  \small
  \setlength{\tabcolsep}{4pt}
  \begin{tabular}{l *{4}{S[table-format=+3.1]} *{4}{S[table-format=<1.2]}}
    \toprule
    & \multicolumn{4}{c}{Runtime reduction (\%)} & \multicolumn{4}{c}{Energy error (\%)} \\
    \cmidrule(lr){2-5}\cmidrule(lr){6-9}
    Approximation & {Test01} & {Test02} & {Test03}& {Test04}& {Test01} & {Test02} & {Test03}&{Test04} \\
    \midrule

    \multicolumn{9}{l}{\textbf{CCSD}} \\
    DLPNO               & 95.8 & 98.7 & 90.4 & 96.4 & 0.05 & 0.06 &  0.26 & 0.07\\
    idea\_06    & 40.6 &  47.6 & -36.2&  -47.0 & 0.20 & 0.21 & < 0.01  & < 0.01\\
    idea\_13    & 22.9   & 23.9 & N/A& 16.9 &  < 0.01 & < 0.01 & N/A& 0.07\\
    idea\_18    & 29.8 & 31.5  & 31.5 & 30.7& < 0.01 & < 0.01 & < 0.01&  < 0.01\\
    \midrule

    \multicolumn{9}{l}{\textbf{CISD}} \\
    DLPNO               & 76.0 & 90.5  & -10.7 & 62.8 & 0.46 & 0.60 & 1.30 & 0.94\\
    idea\_08   & 16.1 & 19.6 & 21.6 & 27.9 & 0.38 & 0.05& 0.15 & 1.22 \\
    idea\_18   & 14.1 & 21.6 & 24.3  & 27.1 & < 0.01 & < 0.01 & < 0.01& < 0.01\\
    idea\_19   & 23.0 & 20.0 & 23.1 & 26.7 & < 0.01 & < 0.01 & < 0.01& < 0.01\\
    \bottomrule
  \end{tabular}
\end{table}

\begin{figure}[htbp]
  \centering
  \includegraphics[width=1.0\linewidth]{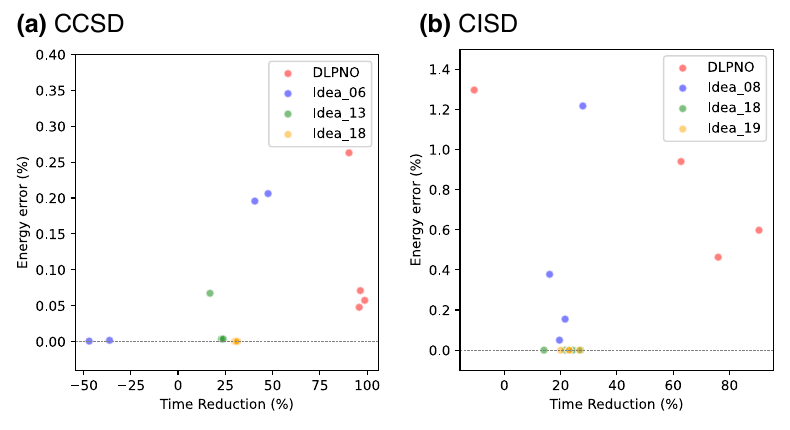}
  \caption{Runtime--accuracy trade-off for DLPNO calculations in ORCA and the top three LADeQ-generated approximation ideas for (a) CCSD and (b) CISD. 
  Each point corresponds to one benchmark molecule. Runtime reduction is reported relative to the unmodified baseline calculation, and correlation-energy error is reported relative to the baseline correlation energy. 
Negative runtime reduction indicates that the approximate calculation is slower than the baseline.}
  \label{fig:speed_error_tradeoff}
\end{figure}

In addition, as a state-of-the-art baseline developed by human experts, we evaluated the established domain-based local pair natural orbital (DLPNO)\cite{Riplinger2013-ur} approach using ORCA\cite{ORCA} with the same molecular geometries and STO-3G basis set.
Because LADeQ-generated implementations and DLPNO were evaluated relative to different unapproximated baselines, PySCF and ORCA, respectively, this comparison should be viewed as a within-framework speed–accuracy assessment rather than a direct comparison of absolute performance.
DLPNO achieved the largest CCSD speedups in our benchmark, exceeding 90\% for all molecules, which reflects the strength of a mature local-correlation approximation. However, these larger speedups were accompanied by larger correlation-energy errors than those of the most accurate LADeQ-generated implementations. For CISD, DLPNO reduced runtime for three of the four molecules but increased runtime for Test03, whereas idea\_18 and idea\_19 achieved moderate runtime reductions with correlation-energy errors below 0.01\% across all tested molecules. The comparison therefore reveals a speed--accuracy trade-off rather than a single uniformly superior method.
This trade-off is visualized in Figure \ref{fig:speed_error_tradeoff}, which plots runtime reduction against correlation-energy error. 
DLPNO tends to fall in a high-speedup region with higher errors, whereas several LLM-generated approximations, notably idea\_18 for CCSD and idea\_19 for CISD, achieve much smaller errors with moderate speedups.

\clearpage
\subsection{Qualitative analysis}
Table \ref{tab:ideas} summarizes the three approximation ideas that achieved the best performance for CCSD and for CISD, respectively, based on the quantitative evaluation. 
Because the ideas proposed by the LLM often consisted of combinations of multiple approximation and acceleration techniques, the table focuses on the key components that are considered to contribute essentially to the runtime reduction, and reports for each technique its originating field and typical use case. 
For example, CCSD idea\_06 includes Krylov acceleration, which accelerates nonlinear fixed-point iterations by extrapolating the next iterate from a short history of recent iterates and corresponds to the approach commonly known as DIIS\cite{Pulay1980-pp} in quantum chemistry.
In addition, CCSD idea\_13 and idea\_18 accelerate matrix–matrix multiplication by hierarchically partitioning matrices and discarding block products based on a norm-based bound with a tunable threshold. 
Conceptually, this is equivalent to Schwarz screening for two-electron integrals\cite{Haser1989-ae,Subotnik2006-cs} in quantum chemistry.
On the other hand, some of the top-ranked ideas for CISD were found to incorporate approximation techniques established in other fields that have not yet been adopted in quantum chemistry. 
Taken together, these results suggest that our proposed pipeline can not only rediscover known techniques but also systematically transfer established approximation techniques from other domains as viable candidates.

\begin{table}[htbp]
\centering
\caption{Idea list of LLM-suggested approximation ideas for CCSD and CISD.
Methods originate in adjacent fields and are presented in terms of the motivation and the transferable mechanism they offer, together with representative references.}
\label{tab:ideas}

\setlength{\tabcolsep}{4pt}

\begin{tabularx}{\linewidth}{@{}
>{\raggedright\arraybackslash}p{0.10\linewidth} 
>{\raggedright\arraybackslash}p{0.15\linewidth}
>{\raggedright\arraybackslash}X                 
>{\centering\arraybackslash}p{1.2cm} @{}}
\toprule
Idea & Field & Technique / Description & Refs \\
\midrule
\multicolumn{4}{@{}l@{}}{\bfseries CCSD}\\
\midrule
idea\_06 & Computational fluid dynamics(CFD) &
\textbf{Nonlinear Krylov acceleration}\par
A derivative-free convergence accelerator for nonlinear fixed-point iterations.
It extrapolates the next iterate from a small history of recent iterates (and/or residuals) by solving a low-dimensional least-squares problem, often yielding faster convergence with minimal overhead.
& \cite{Anderson1965-bs,Oosterlee2006-wh} \\
\midrule
idea\_13, idea\_18 & Numerical linear algebra &
\textbf{SpAMM (Sparse Approximate Matrix Multiply)}\par
A fast approximate matrix--matrix multiplication for matrices with decay (common in linear-scaling SCF, density-matrix methods, and spectral projection).
It represents matrices in a hierarchical block tree and culls block products using a norm-based bound, e.g.\ skipping \(A_{ik}B_{kj}\) when \(\|A_{ik}\|_F\|B_{kj}\|_F \le \tau\), thereby controlling the accuracy via \(\tau\) (recovering the exact product at \(\tau=0\)).
& \cite{Challacombe2010-kg}\\
\midrule

\multicolumn{4}{@{}l@{}}{\bfseries CISD}\\
\midrule
idea\_08 & Kernel methods &
\textbf{Nyström approximation}\par
A low-rank approximation for a large symmetric positive semidefinite (PSD) kernel/covariance matrix \(K\):
\(\tilde K = C W^{\dagger} C^{\mathsf T}\), where \(C\) contains sampled columns of \(K\) and \(W\) is the corresponding sampled submatrix.
It preserves symmetry/PSD by construction and reduces the cost of kernel methods and spectral computations; data-dependent sampling (e.g., leverage-score variants) improves accuracy for a given sample size.

& \cite{Williams2000-en, Drineas2005-iz} \\
\midrule
idea\_18 & Circuit simulation &
\textbf{Constraint-preserving projection-based model order reduction}\par\par
A projection-based reduction for large linear circuit models that enforces stability-related structural constraints during compression.
This prevents unphysical behavior introduced by reduction and enables fast repeated transient/AC simulations in large-scale interconnect analysis.
& \cite{Kerns1997-ee, Odabasioglu1998-rx} \\
\midrule
idea\_19 & Spatial statistics &
\textbf{Circulant embedding}\par
Exploit translation-invariant structure by extending a large correlation/covariance matrix to a periodic form that FFT can diagonalize.
This turns repeated costly matrix operations into fast \(O(n\log n)\) computations, enabling scalable large-system calculations (exact when the extension preserves the required positivity/stability constraint).
& \cite{Wood1994-oc, Dietrich2006-if} \\
\bottomrule
\end{tabularx}

\end{table}

\clearpage
\section{Discussion and perspective}

\subsection{Transferability across molecular classes.}
The cross-molecule benchmarks reveal a systematic pattern that bears on how LADeQ-generated approximations should be deployed in practice.
Approximations that exploit tensor sparsity or spatial locality, mechanisms that arise naturally in saturated, electronically compact systems such as the alkane family, tend to degrade or lose stability when applied to extended $\pi$-conjugated molecules such as coronene.
In such systems, orbital delocalization spreads correlation contributions broadly across the molecular framework and suppresses the sparsity the approximation relies upon.
This is not merely a limitation but an informative signal: the effectiveness of a given approximation is coupled to the electronic structure of the target molecule, and this coupling can be probed empirically within the LADeQ framework itself.
Accordingly, one productive mode of use is not to seek a single universally transferable strategy, but to identify which approximation mechanisms are well-matched to a given molecular class and to flag those requiring recalibration before transfer to chemically distinct systems.
At the same time, transferable strategies are discoverable within the same search.
For CCSD, idea\_18 maintained an approximately 30\% runtime reduction with negligible accuracy loss across all four benchmark molecules, including the delocalized $\pi$ systems.

\subsection{Comparison to established approximation frameworks.}
Comparison with DLPNO, a mature, expert-engineered local-correlation framework, clarifies the nature of the trade-off that LADeQ navigates rather than resolves.
DLPNO represents sustained development within a fixed approximation family, and its ability to deliver end-to-end CCSD speedups exceeding 90\% in our benchmarks reflects the payoff of that investment.
LADeQ, by contrast, is not a refined approximation but a search procedure, and the implementations reported here are early-stage candidates from a single round of exploration.
The more consequential contrast is one of kind rather than degree: DLPNO is a closed-source implementation committed to a predetermined locality principle, whereas LADeQ operates directly within open-source code and is bound to no single mechanism.
This has two practical consequences.
First, LADeQ can draw from a substantially broader design space, as evidenced by the appearance of techniques from spatial statistics, circuit simulation, and kernel methods among its top-ranked discoveries, domains with little established presence in quantum chemistry.
Second, because the generated implementations are fully inspectable and modifiable, they can serve as a starting point for researchers seeking to extend or combine approximation strategies, rather than as an opaque black box.
This adaptability also distinguishes LADeQ from surrogate approaches such as machine-learned interatomic potentials, which deliver speed only after substantial investment in dataset construction, pretraining, and model maintenance.
LADeQ instead constructs approximations at test-time, requiring no pretraining or curated data.

\subsection{Limitations and outlook.}
The present study is deliberately scoped to enable controlled evaluation.
All experiments use the STO-3G basis set, target ground-state correlation energies, and benchmark four molecules in a single-threaded setting.
These choices facilitate rapid iteration and clean comparison, but they also bound the current conclusions.
Establishing the robustness and generality of the approach will require larger and more realistic basis sets, a broader molecular benchmark, and extension beyond CCSD and CISD to multireference methods, periodic codes, and embedding frameworks.
Particularly important are the strongly correlated and open-shell regimes where standard mean-field methods break down and high-level wavefunction treatments become indispensable, since it is precisely there that adaptive, on-demand approximation design stands to be most valuable.
A further direction concerns the scope of acceleration.
The approximations reported here target individual bottlenecks, whereas applying LADeQ iteratively, directing each successive round at the bottlenecks that survive earlier optimizations, offers a path toward end-to-end speedups comparable to those of mature frameworks.
A distinctive feature of building on an out-of-the-box LLM is that LADeQ inherits external progress at no additional cost.
Because the workflow requires no task-specific training, advances in the underlying model translate directly into stronger bottleneck analysis, broader ideation, and more reliable code generation, so the quality of the discovered approximations is expected to improve simply as more capable models become available.
More broadly, because the agent is not constrained by any single researcher's disciplinary background, a compelling open question is whether cross-domain transfer can be pushed beyond the adaptation of existing techniques toward genuinely new approximation mechanisms.

\section{Conclusion}

We introduced LADeQ, an LLM-guided workflow for designing approximations within existing quantum chemistry codes.
Given a baseline electronic-structure implementation, LADeQ prompts an LLM to localize dominant bottlenecks, generate candidate approximation ideas, translate them into concrete code, and iteratively refine each implementation against accuracy and timing feedback until a prescribed error tolerance is met.
Applied to CCSD and CISD, the pipeline autonomously produced multiple implementations that reduced wall-clock time while preserving accuracy, with several approximations transferring across molecules of differing size and electronic character.
A number of the discovered strategies originated in fields outside quantum chemistry, indicating that an LLM agent can surface approximation techniques beyond the reach of any individual researcher's expertise.
While broader validation across larger basis sets, more diverse molecules, and additional electronic-structure methods remains necessary, these results show that LLM-guided approximation design can yield inspectable implementations with measurable speedups and quantifiable accuracy loss.
We view this as an early but concrete step toward agent-driven algorithm discovery in quantum chemistry.

\section{Methods}
\subsection{Overview of the LADeQ workflow}
In this study, we develop LADeQ (LLM-guided Automatic approximation Design for Quantum chemistry), a workflow that uses a large language model (LLM) to automatically design approximations for quantum-chemical codes at both the algorithmic and numerical levels, with the goal of identifying accelerations that remain within a prescribed error tolerance (Figure~\ref{fig:overall_flowchart}).
LADeQ consists of three phases: (i) bottleneck analysis of the baseline code and generation of approximation ideas, (ii) generation of approximation-integrated code together with instrumentation (debugging/trace) mechanisms, and (iii) iterative refinement based on comparisons of error and wall-clock time against the baseline.
We apply LADeQ to coupled-cluster singles and doubles (CCSD) and configuration interaction singles and doubles (CISD), using the corresponding PySCF implementations as baselines, and evaluate the resulting approximation implementations in terms of correlation-energy error and runtime.

\subsection{Target methods and baseline implementations}
The target methods are the reference implementations of CCSD and CISD in PySCF~\cite{PySCF}. 
Hereafter, we refer to the unmodified PySCF implementation as the \emph{baseline} and the modified implementation with an LLM-proposed approximation as the \emph{approximate implementation}.
We verified in advance that the baseline reproduces the reference correlation energy for the test system.
All benchmarks were performed on a linear hydrocarbon molecule, C$_{15}$H$_{32}$ (Figure~\ref{fig:bench_mol}(a)), with the STO-3G basis set.
All calculations were performed on CPU only (Intel Xeon Platinum 8480+) using a single thread.
The LLM used in this study was GPT-5 by OpenAI (accessed Sep. 2025 to Dec. 2025).

\subsection{Phase 1: Bottleneck analysis and approximation-idea generation}
In Phase~1, we aim to generate multiple promising approximation approaches by having the LLM analyze the baseline code and propose acceleration ideas.

\paragraph{Analysis of theory and computational workflow.}
We provided the LLM with the baseline source code and the baseline results for the test molecule (e.g., energies).
The LLM was instructed to analyze the computational workflow of the target theory (CCSD/CISD), identify major tensor operations, determine the dominant terms in computational and memory cost, and localize the dominant bottlenecks (i.e., the kernels/loops that govern the runtime).
\paragraph{Proposal and curation of approximation ideas.}
Based on the analysis above, the LLM proposed approximation ideas aimed at accelerating the dominant bottlenecks.
We curated and recorded each idea as structured text descriptions.
\paragraph{Iterative generation to ensure diversity.}
We fed back the analysis and the accumulated ideas to the LLM and instructed it to generate additional, non-redundant approximation ideas with distinct design principles.
This process was repeated until 20 approximation ideas were obtained for each method.

\subsection{Phase 2: Code generation with approximations and instrumentation}
In Phase~2, the LLM implemented each approximation idea as concrete code modifications to the PySCF baseline and augmented the resulting code with instrumentation to enable diagnosing approximation behavior and performance.

\paragraph{Generation of approximation-integrated implementations.}
Given the baseline code and an approximation idea, the LLM generated a concrete patch that integrates the approximation into the bottleneck region(s) of the baseline implementation.

\paragraph{Automatic insertion of instrumentation.}
To make the impact of the approximation measurable and debuggable (e.g., sources of energy error, convergence behavior, and runtime breakdown), the LLM was instructed to add instrumentation such as logging, lightweight profiling, and intermediate-quantity checks.

\paragraph{Validation of instrumentation adequacy (up to 5 iterations).}
We executed each approximate implementation on the test molecule and examined whether the instrumentation outputs were sufficient to analyze the approximation behavior (e.g., to localize error contributors or identify failure modes).
If not, we returned to the instrumentation step and asked the LLM to augment the debugging/trace code.
This loop was performed up to five times.
To promote implementation diversity, each approximation idea was realized through 10 independent implementation trials. Each trial started from the same baseline code and the same approximation-idea description, but used an independent LLM generation trajectory. These trials were treated as independent attempts to instantiate the same high-level idea in code, yielding 200 implementation trials per target method.

\subsection{Phase 3: Benchmarking and iterative refinement}
In Phase~3, we benchmarked each approximate implementation against the baseline and iteratively refined it until it achieved both acceptable accuracy and a measurable speedup.

\paragraph{Acceptance criteria.}
For each approximate implementation, we compared its output on the test molecule against the baseline in terms of correlation-energy relative error and wall-clock time.
An implementation was accepted if (i) the correlation-energy relative error was within 10\% and (ii) the wall-clock time was at least 2\% shorter than that of the baseline. The 2\% threshold was chosen to exclude timing differences within typical run-to-run variability.

\paragraph{Refinement loop (up to 10 iterations).}
Within each implementation trial, if the acceptance criteria were not met, we provided the LLM with the current code, the computed results, the instrumentation logs, and the timing measurements, and asked it to revise the implementation based on suspected bottlenecks and error sources. We then re-ran the benchmark and repeated the evaluation. This within-trial refinement loop was continued until the criteria were satisfied or a maximum of 10 refinement iterations was reached.

\subsection{Metrics and measurements}
As an accuracy metric, we used the correlation energy relative error defined by
\begin{equation}
\varepsilon = \frac{\left|E_\mathrm{corr}^\mathrm{approx} - E_\mathrm{corr}^\mathrm{base}\right|}{\left|E_\mathrm{corr}^\mathrm{base}\right|},
\end{equation}
where $E_\mathrm{corr}^\mathrm{base}$ and $E_\mathrm{corr}^\mathrm{approx}$ denote the correlation energies obtained from the baseline and approximate implementations, respectively.
The correlation energy was defined as the difference between the total energy and the Hartree--Fock energy for each method.
For runtime, we used wall-clock time and compared the baseline runtime $T^\mathrm{base}$ and the approximate runtime $T^\mathrm{approx}$ under identical conditions.

\section*{Data availability statement}
The data and source code supporting the findings of this study will be made available in a public repository upon publication.

\section*{Competing interests}
The authors declare no competing interests.

\section*{Author contributions}
\textbf{M.H.}: Conceptualization, Methodology, Software, Formal analysis, Investigation, Data curation, Writing -- original draft, Writing -- review \& editing.
\textbf{Y.S.}: Conceptualization, Supervision, Formal analysis, Writing -- review \& editing.
\textbf{T.M.}: Writing -- review \& editing, Formal analysis.
\textbf{S.K.}: Formal analysis, Writing -- review \& editing.
\textbf{M.A.}: Writing -- review \& editing, Formal analysis.

\bibliography{main}
\bibliographystyle{naturemag-doi-eprint.bst}

\clearpage
\appendix

\begin{center}
    \LARGE{\textbf{Appendix}}
\end{center}

\section{Prompts used in LADeQ}
\label{app:prompts}

\begin{tcolorbox}[colback=gray!5,colframe=black!50,title=Prompt for identifying bottleneck of original code in Phase1]
You try to analyze given quantum chemistry calculation code.
First, please specify the theory to be calculated in given code.
Second, please analyze its computational cost in detail from debug log.

Given code:
\{code\}

Debug log:
\{log\}
\end{tcolorbox}

\begin{tcolorbox}[colback=gray!5,colframe=black!50,title=Prompt for suggestion of new approximation method in Phase1]
Your task is to develop a new approximation method for the provided quantum-chemistry computation code.”

The analysis report for the given code is as follows:
\{report\_analysis\}

Here are past suggestions for your reference:
\{past\_suggest\_text\}

Now, please suggest new approximation method which is different from the past suggestions and its cost is less than original cost.
Please propose new approximation method that, to the best of current knowledge, have not been previously reported in the field of computational chemistry.

The given code is here.
{code}

Please suggest one possible method with pseudo code.

Please write the approximation method in `approximation`.
Please write pseudo code in markdown format for `pseudo\_code`.
Please summarize your approximation method in short sentences as well for `summary`.
\end{tcolorbox}

\begin{tcolorbox}[colback=gray!5,colframe=black!50,title=Prompt for implementing approximation method in Phase2]
Your task is to implement the given approximation method into the given original code.
    
original code:
\{original\_code\}
    
approximation method:
\{approximation\_method\}
    
Set the approximation as the default behavior in the implementation.
You have to add debug mode to observe approximation effects by setting a flag `debug` in the `run` function.
You also have to add approximation mode to observe approximation effects by setting a flag `approximate` in the `run` function.
Set the approximation as the default behavior in the implementation.
Please leave IO handling parts as it is, and only implement the core part of the approximation method.
\{DIFF\_FORMAT\_INSTRUCTION\}
Describe diff code to the original code in `diff`.
You also need to explain how your implementation works in `explanation`.
\end{tcolorbox}

\begin{tcolorbox}[colback=gray!5,colframe=black!50,title=Prompt for judging whether debug logs are sufficient in Phase2]
Your task is judge whether the debug logs are sufficient to verify the performance of the approximation method.
If you judge the debug logs are sufficient, respond with `is\_debug\_print\_enough` as True and provide an analysis in `analysis`.
    
You will be given the current runtime logs, both with and without debug mode enabled and both with and without the approximation mode enabled.
    
\{runtime\_logs\}

\{runtime\_errors\}

You try to implement the approximation method into the original code.
The above logs and errors are the results of the implementation.
You need to make diff code by adding debug function to examine whether the approximation method is functioning correctly.

The current implementation code is as follows:
\{impl\_code\}

approximation method:
\{approximation\_method\}

If the above debug logs are not sufficient, your task is to add debug function to examine whether the approximation method is functioning correctly.

Do not determine whether the approximation method is effective in this task. Instead, add as many debug prints as reasonably possible to help verify the effectiveness of the approximation method. Avoid producing overly large logs--do not dump entire matrices or similarly large data structures. Examples of debug information to add are listed below.
- Execution-time profile per processing block
- Number of iterations (loops) to reach convergence
- Matrix sizes
Additionally, perform any supplementary profiling and print-based debugging appropriate to the characteristics of the approximation.

Please take care that you can change pyscf logger level by changing verbose such as verbose=logger.DEBUG1.

\{DIFF\_FORMAT\_INSTRUCTION\}
Describe diff code to add the debug function to examine whether the approximation method is functioning correctly in `updated\_diff`.
If the debug logs are sufficient to verify the performance of the approximation method, respond with `is\_debug\_print\_enough` as True and provide an analysis in `analysis`.
Describe why you made those changes in `analysis`.
\end{tcolorbox}

\begin{tcolorbox}[colback=gray!5,colframe=black!50,title=Prompt for refining approximation in Phase3]
Your task is to judge whether the approximation method was used effectively based on the debug logs by comparing the results with and without the approximation method.
If not, update the diff code to make the approximation method effective.

The permitted error threshold brought by the approximation method is 10\% of the correlated energy value.
If the total calculation time with the approximation method is reduced compared to that without the approximation method, and the error is within the permitted threshold, you can consider that the approximation method was used effectively.
In addition, the approximation method will be regarded as effective only if a speedup of at least 2\% is observed.

You must not change Davidson’s tolerance or conv\_tol or conv\_tol\_normt or max\_iter from their default values.

\{runtime\_logs\}

\{runtime\_errors\}

The current implementation code is as follows:
\{impl\_code\}

approximation method:
\{approximation\_method\}

\{DIFF\_FORMAT\_INSTRUCTION\}

Describe diff code to make the approximation method effective in `updated\_diff`.
If the approximation method was used effectively based on the debug logs, respond with `is\_approximation\_enough` as True.
You also need to provide an analysis in `analysis`.
\end{tcolorbox}

\clearpage
\section{Implementation-level speed--accuracy distributions}
\label{app:speed-error-scatter}

\begin{figure}[htbp]
  \centering

  \begin{subfigure}{0.7\linewidth}
    \centering
    \begin{overpic}[width=\linewidth]{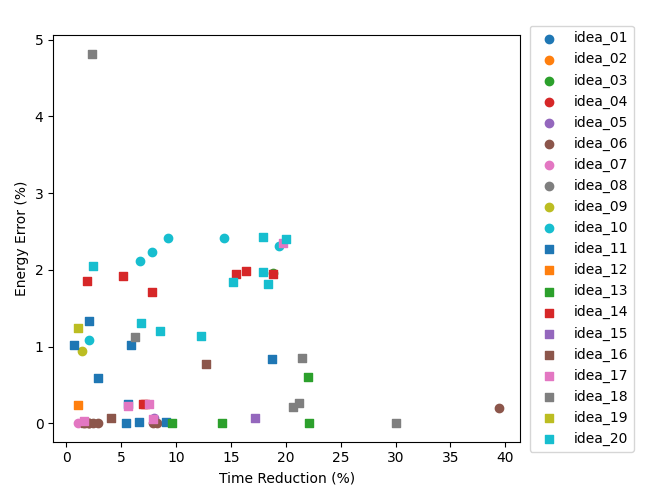}
      \put(0,70){\large\bfseries (a)}
    \end{overpic}
    \label{fig:ccsd_time_error_relation}
  \end{subfigure}

  \vspace{1em}

  \begin{subfigure}{0.7\linewidth}
    \centering
    \begin{overpic}[width=\linewidth]{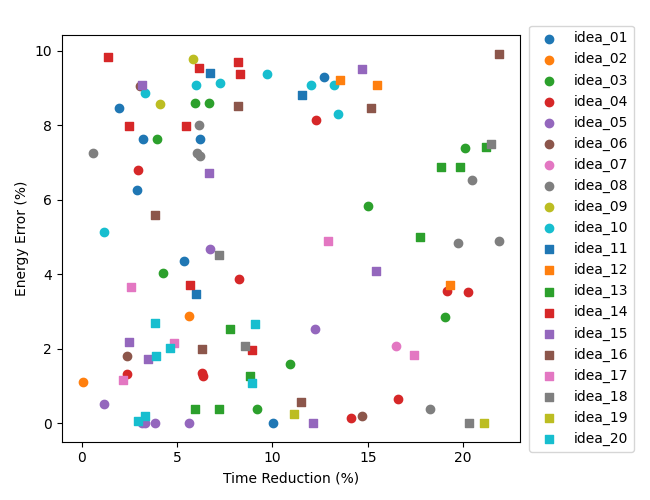}
      \put(0,70){\large\bfseries (b)}
    \end{overpic}
    \label{fig:cisd_time_error_relation}
  \end{subfigure}

    \caption{Implementation-level speed--accuracy distributions for the LLM-generated approximation candidates for (a) CCSD and (b) CISD.
    Each point corresponds to one LLM-generated approximate implementation.
    The horizontal axis shows the runtime reduction relative to the unmodified baseline, and the vertical axis shows the relative correlation-energy error with respect to the baseline correlation energy.
    Colors indicate approximation ideas.
    The lower-right region corresponds to desirable implementations with larger runtime reductions and smaller energy errors.}
  \label{fig:time_error_relation}
\end{figure}

Figure~\ref{fig:time_error_relation} shows the implementation-level distribution of runtime reduction and correlation-energy error for the LLM-generated approximation candidates.
While the main text reports the best-performing implementation for each idea under the prescribed error constraint, these scatter plots show the broader distribution across generated implementations.
Each point corresponds to one approximate implementation, with colors indicating the approximation idea.
The lower-right region corresponds to desirable implementations, namely those that reduce runtime while maintaining small correlation-energy error.
These plots illustrate that implementation outcomes vary substantially even for the same high-level idea, motivating the use of multiple independent implementation trials per idea.

\end{document}